\documentclass[np2,twoside,onecolumn,showpacs]{revtex4}

\usepackage{amssymb}
\usepackage{amsmath}
\usepackage{graphicx}
\usepackage{bm,multirow}

\begin{document}
 
%\setcounter{page}{1}

%\title[$\ll$ Review Article ]{From Resistance Minimum to Kondo Physics}
\title{From Resistance Minimum to Kondo Physics}

\author{Takashi \surname{YANAGISAWA}}
\email{t-yanagisawa@aist.go.jp}
\affiliation{National Institute of Advanced Industrial Science and Technology,
1-1-1 Umezono, Tsukuba, Ibaraki 305-8568, Japan}

\date[]{Submitted 30 September 2023}

\begin{abstract}
We discuss the development of Kondo physics from the resolution of the resistance
minimum by J. Kondo to recent developments in physics.
The Kondo effect has given a great impact to all areas of physics.
This reminds us that physics is one unified science.  Kondo's pioneering work has led
major developments in physics.  We show brief history of the Kondo effect and discuss the
Kondo effect from several points of view that appeared to be important through
conversations with Prof. J. Kondo.  
\end{abstract}

\pacs{72.15.Qm, 75.30.Mb\\Keywords: Kondo effect, resistance minimum, Kondo problem, Kondo physics}

\maketitle

\section{Brief history of Prof. Kondo}

Prof. Jun Kondo was born in 1930 in Tokyo and is well known for explaining the
strange behavior of electrical resistivity of some metals at low temperatures called the
resistance minimum. This kind of anomalies mainly arises in metals called the dilute
magnetic alloys.  In 1964 Kondo at the Electrotechnical Laboratory solved this
mystery based on the s-d model by showing that electron scattering is strongly
enhanced at low temperature and the resistivity increases\cite{kon64}.  

In 1954 he graduated from the University of Tokyo and began his research career in
the graduate school at Institute of Science and Technology in Komaba where his
supervisor was Prof. T. Muto, and Prof. J. Yamashita was the asociate professor
in this laboratory.
Kondo's Doctor thesis was on superexchange interaction in oxides such as 
MnO\cite{kon57,kon58,kon59,kon59b}.
In 1960 he was at the Institute for Solid State Physics as a research associate and
studied magnetism in metals and anomalous Hall effect of ferromagnetic metals\cite{kon62}.
In this study he investigated the s-d model with skew scattering showing that
the Hall conductivity is proportional to $\langle (M-\langle M\rangle)^3\rangle$.

In 1963 he moved to the Electrotechnical Laboratory and started the research on
the resistance minimum.  He succeeded to explain the singular behavior of the
electrical resistivity in dilute magnetic alloys.
Various anomalies that occur in low temperatures are collectively called
the Kondo effect.
The physics of the Kondo effect has developed beyond our expectations after
the Kondo's pioneering work.

He was the fellow of Electrotechnical Laboratory and
became the fellow emeritus after retirement in 1990.
He was then a professor at Toho University from 1990 to 1995.
He was awarded the Fritz London Memorial Award in 1987.  He became the member of the
Japan Academy in 1997 and the member of Foreign Associate of National Academy
of Sciences (USA) in 2009.  In 2020 he received the Order of Culture of Japan.

\section{Resistance Minimum and the Kondo Effect}
  
\subsection{Resistance minimum}

The resistance minimum was found experimentally in 1930s\cite{mei30,haa33}.   
The first paper was published in 1930, where Meissner and Voigt measured the
resistivity for several metals down to about 1.2K and found that the resistivity
at lowest temperature is larger than those at higher temperatures\cite{mei30}.
In 1933 van den Berg et al. found a minimum in the resistivity curve as a
function of temperature\cite{haa33}.
This was about 20 years after the discovery of superconductivity by
H. K. Onnes\cite{onn11}.  The resistance minimum had been regarded as one of the two most
difficult problems in the condensed matter physics in the world.
After this work there appeared many experimental works and this phenomenon
has been observed for many metals.\cite{ger52,kor53,owe56}  
There had been intensive experiments on, for example, dilute alloys of Mn in Cu.
It had been suggested that this phenomenon 
originated from magnetic impurities included in metals.
The s-d model was introduced in 1950s, and almost all experiments other than the resistance
minimum could be understood based on the s-d model at that time.
Kondo examined the review paper by van den Berg\cite{ber62} in detail and also
the Thesis of Dr. Knook at the Leiden University\cite{kno62}. 
He was strongly convinced that there was deep physics behind this phenomenon. 
In particular, the paper by Sarachik et al.\cite{sar64} clearly showed that
the resistance minimum appeared when the impurity had the magnetic moment.

It had been confirmed that the resistance minimum phenomenon is proportional to
the concentration $c$ of magnetic impurities up to about 
1960\cite{ber62,kno62,sar64,mac62,ber64,ber64b}.
This is an important experimental result and it is also significant that the
temperature $T_{min}$ at which the resistance minimum occurs is proportional to 
$c^{1/5}$: $T_{min}\propto c^{1/5}$.
From these two properties the resistivity $R(T)$ can be written as
\begin{equation}
R(T)= aT^5+c\rho_0-c\rho_1 g(T)
\end{equation}
where the first term indicates the lattice resistivity, and the second term arises
from the impurity potential and the spin scattering giving a constant.
The last term, proportional to the impurity concentration $c$, represents the
anomalous term which would appear due to some mechanism.  The temperature $T_{min}$ at which
the resistance minimum is observed is given by the solution of the equation
\begin{equation}
\frac{dR(T)}{dT}= 5aT^4-c\rho_1\frac{dg(T)}{dT}=0.
\end{equation}
The fact that $T_{min}\propto c^{1/5}$ gives a constraint that $dg(T)/dT\propto 1/T$.
This means that $g(T)$ should be the logarithmic function $g(T)=\log T$
and the minimum is given by
\begin{equation}
T_{min}= \left(\frac{c\rho_1}{5a} \right)^{1/5}.
\end{equation}
Thus the experiments had already suggested the existence of the $\log T$ in the resistivity. 
Jun Kondo, like other researchers, of course did not notice the
logarithmic correction at that time.

\subsection{The s-d model}

Prof. Kondo concluded that we should examine the s-d model 
to clarify the origin of the resistance minimum from the experimental results. 
The s-d model is given as\cite{kon64,kon12}
\begin{equation}
H= \sum_{{\bf k}\sigma}\epsilon_{{\bf k}}c^{\dag}_{{\bf k}\sigma}c_{{\bf k}\sigma}
-\frac{J}{N}\sum_{{\bf k}}\big[(c_{{\bf k}\uparrow}^{\dag}c_{{\bf k}\uparrow}
-c_{{\bf k}\downarrow}^{\dag}c_{{\bf k}\downarrow})S_z
+c_{{\bf k}\uparrow}^{\dag}c_{{\bf k}\downarrow}S_{-}  
+c_{{\bf k}\downarrow}^{\dag}c_{{\bf k}\uparrow}S_{+}  \big],
\end{equation}
where $J$ denotes the magnitude of the exchange interaction between the localized
and conduction electrons and $N$ is the total number of atoms in the crystal.
This model was used to examine magnetic interactions of magnetic metallic 
compounds\cite{kas56,yos57}.
This model had already been used in examining the resistance minimum behavior
in dilute magnetic alloys.  It gives the relaxation time given as
\begin{equation}
1/\tau (\epsilon_{{\bf k}})= 2\pi cJ^2S(S+1)\rho(\epsilon_{{\bf k}})/3\hbar,
\end{equation}
where $S$ is the magnitude of the localized spin and $\rho(\epsilon)$
is the density of states of conduction electrons.
This leads to a constant resistivity and cannot explain the resistance minimum.
What was missing?  Kondo tried to calculate higher order contributions and
found that the logarithmic term would appear in the transition probability and
thus in the resistivity.
The important contributions come from processes where spin exchange interaction
occurs.  We have two matrix elements for the scattering 
${\bf k}\uparrow\rightarrow {\bf k}'\uparrow$ as
\begin{equation}
\left(-\frac{J}{N}\right)^2(S_z^2+S_{-}S_{+})\sum_{{\bf k}''}
\frac{1-f_{{\bf k}''}}{\epsilon_{{\bf k}}-\epsilon_{{\bf k}''}}
-\left(-\frac{J}{N}\right)^2(S_z^2+S_+S_-)\sum_{{\bf k}''}
\frac{f_{{\bf k}''}}{\epsilon_{{\bf k}'}-\epsilon_{{\bf k}''}}.
\end{equation}
This gives
\begin{equation}
\left(\frac{J}{N}\right)^2\bigg[ S_z^2+\frac{1}{2}\left(S_+S_{-}+S_-S_+\right)\bigg]
\sum_{{\bf k}''}\frac{1}{\epsilon_{{\bf k}}-\epsilon_{{\bf k}''}}
+\left(\frac{J}{N}\right)^2\left( S_+S_{-}-S_+S_-\right)\sum_{{\bf k}''}
\left( f_{{\bf k}''}-\frac{1}{2}\right)\frac{1}{\epsilon_{{\bf k}''}-\epsilon_{{\bf k}}}
\end{equation}
where we have used the energy conservation $\epsilon_{{\bf k}}=\epsilon_{{\bf k}'}$.
The second term, coming from the commutator $[S_+, S_-]$, contains the important integral 
given by
\begin{equation}
g(\epsilon)= \frac{1}{N}\sum_{{\bf k}''}\left( f_{{\bf k}''}-\frac{1}{2}\right)
\frac{1}{\epsilon_{{\bf k}''}-\epsilon_{{\bf k}}}
=\int \left(f(\epsilon')-\frac{1}{2}\right)\frac{\rho(\epsilon')}{\epsilon'-\epsilon}
d\epsilon'.
\end{equation}
This gives a logarithmic divergence $g(\epsilon)=\rho \log|\epsilon/D|$ for $\epsilon\ll D$
at $T=0$ where we use the constant density of states $\rho(\epsilon)={\rm const.}$
for $-D\le \epsilon\le D$.  This results in the logarithmic term in the resistivity as
\begin{equation}
R= R_0\bigg[ 1+4J\rho\log\left(\frac{k_BT}{D}\right)\bigg].
\end{equation}
This formula well explained the anomalous behavior of the resistivity for $J<0$.
The typical behavior of the resistivity is shown in Fig. 1 where the clear logarithmic
dependence is observed.

\begin{figure}[ht]
\includegraphics[width=7cm]{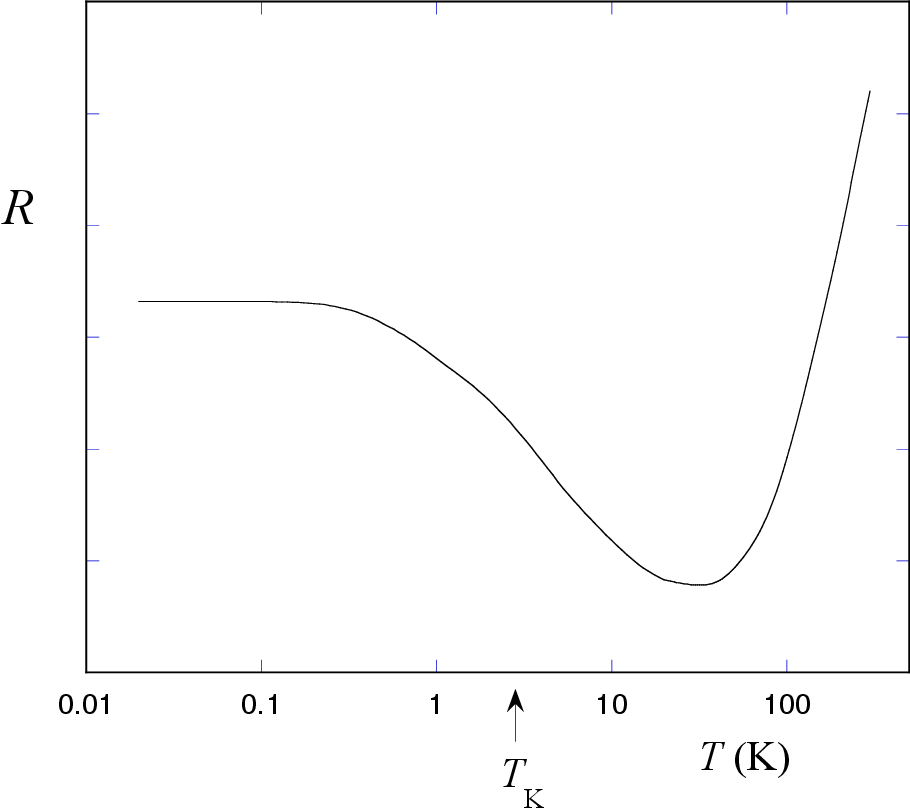} 
\caption{A typical behavior of the resistivity $R$ as a function of the
temperature $T$.  The arrow indicates the Kondo temperature $T_K$.}
\label{fig1}
\end{figure}

\subsection{Kondo problem}

The Kondo theory, however, raised a new problem although the resistance minimum was
clearly explained.  This problem was that $\log T$ has a singularity at absolute zero
$T=0$ and how the physical quantities are expressed as a function of $T$.
This was called the Kondo problem.
The $\log T$ singularity appears in many physical quantities: (1) the spin
relaxation time $(T_1T)^{-1}\propto 1+4J\rho\log(k_BT/D)$, (2) the spin susceptibility
$\chi= \chi_{Curie}\left( 1+2J\rho /[1-2J\rho\log(k_BT/D)] \right)$,
(3) the specific heat\cite{kon68} 
$\Delta c= 16\pi^2 S(S+1)J^4\rho^4 k_B/[1-2J\rho\log(k_BT/D)]^4$,
(4) the entropy $\Delta S= (8\pi^2/3)S(S+1)J^3\rho^3k_B/[1-2J\rho\log(k_BT/D)]^3$.

Kondo examined higher order corrections in the perturbative expansion in terms of
$J\rho$ and found that the effective expansion parameter is $J\rho\log(k_BT/D)$
in stead of $J\rho$\cite{kon69}.  When we consider only the most divergent terms, the physical
quantity $Q(T)$ behaves as
\begin{equation}
Q(T)\propto J^n/\alpha(T)^n ~~~~ {\rm for} ~~ \alpha(T)=1-2J\rho\log(k_BT/D).
\end{equation} 
We have $n=2$ for the resistivity, $n=1$ for the spin susceptibility, $n=3$ for the
entropy and $n=4$ for the specific heat.
For $J<0$, $Q(T)$ diverges at the Kondo temperature $T_K$:
\begin{equation}
k_BT_K = De^{1/2J\rho}.
\end{equation} 

Many theoretical works were done on the Kondo problem during this period called the
Wilderness age.  Abrikosov proposed the method of systematic perturbative expansion
employing quasi-fermion operators and found that the correction to the resistivity
is given by $[1-2J\rho\log(k_BT/D)]^{-2}$\cite{abr65}. 
Suhl applied the Chew-Low method of meson scattering to the s-d model and derived
the equation for the scattering matrix\cite{suh65,suh67}.
Nagaoka used the Green function method for the s-d model to obtain a closed set of
equations based on the decoupling procedure\cite{nag65,ham67,fal67,blo67}.
The solutions by Suhl and Nagaoka turned out to be equivalent later.
Zittarz and M\"{u}ller-Hartmann solved the Nagaoka equation analytically and
obtained an analytic form of the entropy.\cite{zit68}
Although these theories were reliable above $T_K$, 
the low temperature property of the s-d model was still unclear because of
approximations used in these calculations.

The ground state of the s-d model was investigated by using variational wave
functions\cite{yos66,kon66}. 
Yosida proposed a singlet wave function formed by the localized spin and conduction
electrons.  This type of wave function turned out to be correct after the resolution
of the Kondo problem.

\subsection{The Spin fluctuation}

The essential point of the Kondo problem is the spin fluctuation of the magnetic impurity.
The spin of the localized impurity changes its direction with the relaxation rate
$\hbar W$.  When $\hbar W$ is large, up and down spins exist equally and the averaged
value may vanish.  When $\hbar W$ is small, the up spin (or down spin) may exist
giving the Curie type susceptibility.  In the former case we have the Pauli
susceptibility.  The problem is that what should we compare the spin relaxation
rate $\hbar W$ with?  The answer is the temperature $T$.  This is related to the
uncertainty principle.  As shown in Fig. 2, the Korringa relaxation rate
$\hbar W_{ko}$ is small compared to $k_B T$ at high temperature where the localized
spin is in the up or down spin state and may change spin state by scattering
of conduction electrons.  This brings about $\log T$ corrections and this effect
increases as the temperature decreases.
At low temperatures the relaxation rate $\hbar W$ becomes greater than $k_BT$
where the localized spin has up and down spins equally.  In this state
the resistivity increases according to the Friedel sum rule and at the same time the
susceptibility reduces to the Pauli-type susceptibility.\cite{fri52,fri54,fri58}
The state with maximum resistivity at absolute zero is called the unitary limit
(or unitarity limit).
This picture shows that the Kondo effect occurs as a crossover from weakly
correlated region to strongly correlated region and that the logarithmic
correction is a singularity associated with this crossover.

\begin{figure}[ht]
\includegraphics[width=6.5cm]{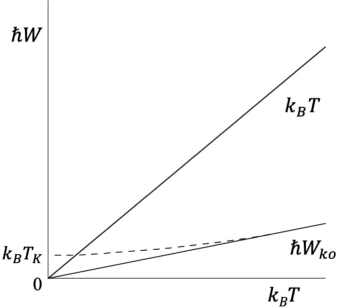} 
\caption{The spin fluctuation rate $\hbar W$ as a function of the temperature $T$.
$\hbar W_{ko}$ indicates the Korringa relaxation rate $\hbar W_{ko}=2\pi(J\rho)^2k_BT$.
The dashed line represents the correct behavior of $\hbar W$.}
\label{fig2}
\end{figure}

\subsection{Renormalization group theory}

The above picture was indeed confirmed by the renormalization group method.
Anderson proposed the scaling equation for the s-d model which he called the
poor man's scaling.\cite{and70}
Anderson, Yuval and Hamann transformed the s-d model to the one-dimensional
classical statistical model interacting through logarithmic potentials (two-dimensional
classical Coulomb gas).\cite{and69,yuv70,and70b}
They derived the renormalization group equations from the Coulomb gas model.
The same equations were also derived\cite{abr70,fow71} in the formulation developed 
by the Gell-Mann and Low.\cite{gel54}
In this subsection we write the s-d interaction in the form (apart from Kondo's notation)
\begin{equation}
H_{s-d}= 
\frac{1}{2N}\sum_{{\bf k}}\big[J_z(c_{{\bf k}\uparrow}^{\dag}c_{{\bf k}\uparrow}
-c_{{\bf k}\downarrow}^{\dag}c_{{\bf k}\downarrow})S_z
+J_{\pm}\left( c_{{\bf k}\uparrow}^{\dag}c_{{\bf k}\downarrow}S_{-}  
+c_{{\bf k}\downarrow}^{\dag}c_{{\bf k}\uparrow}S_{+}\right)  \big],
\end{equation}
where we introduced the anisotropic couplings and
positive $J_z$ and $J_{\pm}$ indicate antiferromagnetic interactions.
Then the renormalization group equations for the exchange coupling are written as
\begin{equation}
\mu\frac{dJ_{\pm}}{d\mu}= -\rho J_zJ_{\pm}, ~~~~  \mu\frac{dJ_z}{d\mu}= -\rho J_{\pm}^2,
\end{equation}
where $J_z$ and $J_{\pm}(=J_x=J_y)$ are exchange coupling constants where we
take account of the anisotropy of $J$ and $\mu$ 
indicates the energy scale.  This results in the important relation
\begin{equation}
J_z^2-J_{\pm}^2= {\rm const}.
\end{equation}
The equations show that $J_z$ and $J_{\pm}$ decrease as the energy scale $\mu$ increases.
This indicates that the s-d model exhibits asymptotic freedom.
The renormalization group flow is shown in Fig. 3.
In the isotropic case, $J\equiv J_z= J_{\pm}$ satisfies
$\mu dJ/d\mu = -\rho J^2$.  We set the initial value of $J$ as $J=J_0$ at $\mu=\mu_0$, and
we have $1/\rho J_0-1/\rho J(\mu)= \ln(\mu_0/\mu)$.  When we suppose that $\rho J(\mu)$
becomes of order one at $\mu=\mu_K$, $\mu_K$ is given by the Kondo temperature for
$\mu_0=D$:
\begin{equation}
\mu_K \simeq \mu_0\exp\left( -\frac{1}{\rho J_0} \right) = D\exp\left( -\frac{1}{\rho J_0} \right)
= T_K.
\end{equation}

\begin{figure}[ht]
\includegraphics[width=7.0cm]{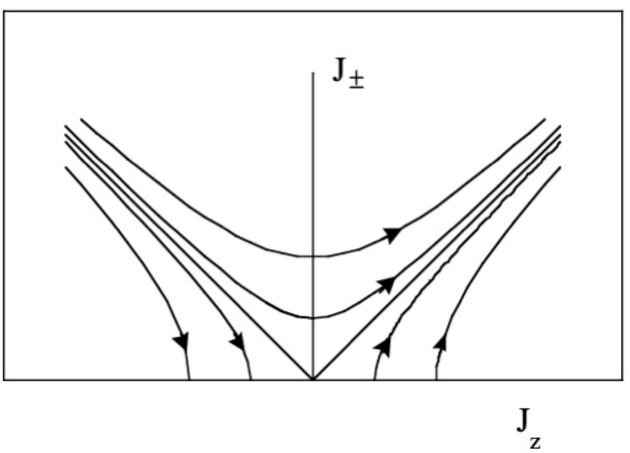} 
\caption{Renormalization group flow in the $J_z-J_{\pm}$ plane.
The scaling is toward the strong coupling region as the energy scale is reduced
$(\mu\rightarrow 0)$ for the antiferromagnetic couplings.
}
\label{fig3}
\end{figure}

The s-d model is closely related to other theoretical models.
We set $x=-J_z\rho$ and $y= J_{\pm}\rho$.
Then we obtain
\begin{equation}
\mu\frac{\partial x}{\partial\mu}= y^2, ~~~~ \mu\frac{\partial y}{\partial\mu}= xy.
\end{equation}
These are equations for the Kosterlitz-Thouless transition,\cite{ber72,kos73,kos74} and we have
$x^2-y^2={\rm const}$.

Since the two-dimensional classical Coulomb gas model is mapped to the sine-Gordon model,
the physical state of the s-d model corresponds to the phase of asymptotic freedom in 
the sine-Gordon model.
We write the sine-Gordon model in the form
\begin{equation}
S_{SG}= \int d^2x \bigg[ \frac{1}{2t}(\nabla\phi)^2+\frac{\alpha}{t}\cos(\phi) \bigg],
\end{equation}
for the scalar field $\phi$.  $t$ and $\alpha$ are coupling constants that will be
renormalized to compensate divergences. 
The partition function can be expanded in terms of $g\equiv \alpha /t$ as\cite{zin93}  
\begin{equation}
Z_{SG}= \sum_{n=0}^{\infty}\frac{1}{(n!)^2}\left(\frac{g}{2}\right)^2\int d^2x_1\cdots
d^2x_n d^2y_1\cdots d^2y_n \exp\bigg[ \frac{t}{2}\left( \sum_{i\neq j}G(x_i,x_j)
+\sum_{i\neq j}G(y_i,y_j)-2\sum_{ij}G(x_i,y_j) \right)\bigg],
\end{equation}
where $x_i$ and $y_i$ are two-dimensional coordinates and $G(x,y)$ is the
two-dimensional Green function
\begin{equation}
G(x,y)= \frac{1}{2\pi}\ln|x-y|.
\end{equation}
For the s-d model, the effective action is given by $(2-\epsilon)V$\cite{and70b} where
$\epsilon= 2J_z\tau$ and
\begin{equation}
V= \sum_{i>j}\left( \ln\frac{|x_i-x_j|}{\tau}+\ln\frac{|y_i-y_j|}{\tau} \right)
-\sum_{i,j}\ln\frac{|x_i-y_j|}{\tau},
\end{equation}
where $\tau$ is the small cutoff.
The factor 2 in $(2-\epsilon)V$ comes from spin degrees of freedom.
In order to make the correspondence with the sine-Gordon model and to perform the
renormalization procedure,
it is preferable that this factor is 4.
Thus we double the effective action (by introducing (orbital) degeneracy for both the conduction
and localized electrons) and use the following correspondence
\begin{equation}
t= 4\pi (2-2J_z\tau), ~~~~ g= \alpha/t = J_{\pm}\tau.
\end{equation}
The renormalization group equations for the sine-Gordon model in two dimensions 
in the lowest order theory are 
given by\cite{ami80,yan16,yan21,yan17}
\begin{equation}
\mu\frac{\partial\alpha}{\partial\mu} = -\left(2-\frac{t}{4\pi}\right)\alpha, ~~~~~
\mu\frac{\partial t}{\partial\mu} = \frac{1}{32}t\alpha^2.
\end{equation}
Here the coefficient $1/32$ in the second equation may depend on the renormalization
procedure, and the Wilson renormalization method\cite{wil74} gives a different coefficient.
From this set of equations we obtain 
\begin{equation}
\mu\frac{\partial J_{\pm}}{\partial\mu}= -2\rho J_zJ_{\pm}, ~~~~~ 
\mu\frac{\partial J_z}{\partial \mu}= -2\pi^2\rho J_{\pm}^2,
\end{equation}
where we set $\rho=\tau$.  
The equations agree with those for the s-d model except for an extra factor $\pi^2$
in the second equation,
which may reflect the uncertainty of coefficients in renormalization group equations
for the sine-Gordon model.

Therefore the Kondo system (s-d model) is in a universality class which contains the
sine-Gordon model and the Kosterlitz-Thouless transition, and is the most
important system in this class.

\subsection{Numerical renormalization group method}

K. G. Wilson succeeded to calculate physical quantities down to absolute zero $(T=0)$
by employing the numerical renormalization group method.\cite{wil75}
Wilson calculated the renormalization group equation exactly employing a numerical
method, and showed that the scaling curve flows to the strong coupling limit.
Since the localized electron interacts with the $s$-wave component of conduction
electrons, the s-d model is reduced to a one-dimensional model.
He considered the Hamiltonian given as
\begin{equation}
H_K= \sum_{\sigma}\int_{-1}^{1}dk a_{k\sigma}^{\dag}A_{k\sigma}
-J(A^{\dag}\vec{\sigma} A)\cdot\vec{\tau},
\end{equation}
where 
\begin{equation}
A_{\sigma}=\int_{-1}^1 a_{k\sigma}dk.
\end{equation}
$a_{k\sigma}$ should satisfy $\{a_{k\sigma}^{\dag}, a_{k'\sigma}\}=\delta(k-k')$.
In order to apply the numerical renormalization group method, $H_K$ was written in the form
\begin{equation}
H = \sum_{\sigma}\sum_{n=0}^{\infty}\Lambda^{-n/2}(f_{n\sigma}^{\dag}f_{n+1\sigma}
+f_{n+1\sigma}^{\dag}f_{n\sigma})
-\tilde{J}(f_0^{\dag}\vec{\sigma}f_0)\cdot\vec{\tau},
\end{equation}
where $f_{n\sigma}$ are a set of discrete electron destruction operators. 

Wilson found that the ground state is a spin singlet formed between the localized spin 
and conduction electrons 
and that the Fermi liquid state is realized at
low temperatures.
The s-d model exhibits a typical system where the Fermi liquid state is realized.

\subsection{Fermi liquid state}

In general the Fermi liquid state can be described by a small number of parameters.
Wilson found an important fact that the low energy properties of the s-d model are 
described by 
two parameters at low temperatures.
We can choose two parameters, for example, the specific heat coefficient $\gamma$
and the spin susceptibility $\chi$ of the localized spin.
At $T=0$ the susceptibility is given by\cite{kon12,and83}
\begin{equation}
\chi = \frac{\mu_B^2}{\pi T_0},
\end{equation}
and the specific heat coming from the localized spin ($k_B=1$) is
\begin{equation}
C= \frac{\pi T}{6T_0},
\end{equation}
where
\begin{equation}
T_0= \sqrt{\frac{e}{\pi}}\frac{1}{1.2002}T_K.
\end{equation}
These results were obtained by applying the Bethe ansatz to the 
s-d Hamiltonian.\cite{and83,and80,and81,wie81,fil81,tsv83}
Wilson found numerically that
\begin{equation}
R_W \equiv \frac{\chi T}{C}\frac{\pi^2}{3}\frac{k_B^2}{\mu_B^2}= 2.
\end{equation}
Because of this relation, we have only one parameter to describe low temperature
properties.  This one parameter is nothing but the Kondo temperature $T_K$.

The phase shift $\delta$ is also important.  When we suppose that the conduction
electron wave function is given by $\cos(x)$ in one dimensional space and that there
is a localized spin at $x=0$, then the wave function should be $\sin(x)$ since
the conduction electrons and localized spin form a singlet at $x=0$ and the conduction
wave function should vanish there.  This indicates that we have the phase shift
$\delta= \pi/2$ at $T=0$.  The phase shit $\delta$ at the Fermi surface 
indicates the
number of conduction electrons which are localized around the localized spin
according to the Friedel sum rule. 
When we adopt 
$\delta(\epsilon)=\delta_0+\epsilon\alpha+\cdots$ for the energy $\epsilon$ measured
from the Fermi surface, the relaxation time $\tau(\epsilon)$ is given by
\begin{equation}
\frac{1}{\tau(\epsilon)}= \frac{2}{\hbar\pi\rho_0}\sin^2\delta(\epsilon)
\simeq \frac{1}{\hbar\pi\rho_0}(1-\alpha^2\epsilon^2+\cdots).
\end{equation}
This explains the $T^2$ dependence of the resistivity. 
Nozi\'{e}res proposed the following form of the phase shift for the spin $\sigma$:\cite{noz74}
\begin{equation}
\delta_{\sigma}(\epsilon) = \delta_0+\alpha\epsilon+\sigma\phi^a m,
\end{equation}
where $m=n_{\uparrow}-n_{\downarrow}$.  When we regard $\delta_{\sigma}/\pi\rho_0$ as 
the shift of electron energy, $\epsilon_{\sigma}$ shifts to 
\begin{equation}
\epsilon_{\sigma}-\frac{\alpha}{\pi\rho_0}\epsilon_{\sigma}-\sigma\frac{\phi^a}{\pi\rho_0}m
-\sigma\frac{1}{2}g\mu_BH,
\label{energy}
\end{equation}
where the Zeeman term is included.  $1-\alpha/\pi\rho_0$ indicates the mass
enhancement factor and the contribution to the specific heat from the localized spin is
$\delta\rho= \rho_0\left((1-\alpha/\pi\rho_0)^{-1}-1\right)=(\alpha/\pi)(1-\alpha/\pi\rho_0)^{-1}$.
The energy in eq.(\ref{energy}) vanishes on the Fermi surface, which indicates
$\epsilon_{\sigma}(1-\alpha/\pi\rho_0)=\sigma(\phi^am/\pi\rho_0+g\mu_BH/2)$.
The magnetization is obtained from $n_{\sigma}=\rho_0(\epsilon_{\sigma}+D)$ and this leads
to the susceptibility $\chi= (1/2)g\mu_Bm/H$.  The localized-spin part of $\chi$ is
$\chi_d=\chi-(1/2)(g\mu_B)^2\rho_0$ given by
\begin{equation}
\chi_d= \frac{1}{2}(g\mu_B)^2\rho_0\left( \frac{\alpha}{\pi\rho_0}+\frac{2\phi^a}{\pi}\right)
\frac{1}{1-\frac{\alpha}{\pi\rho_0}-\frac{2\phi^a}{\pi}}
\simeq \frac{1}{2}(g\mu_B)^2\rho_0\left( \frac{\alpha}{\pi\rho_0}+\frac{2\phi^a}{\pi}\right),
\end{equation}
where $\alpha/\rho_0$ and $\phi^a$ are assumed to be small.
Since the Wilson ratio should be 2,
\begin{equation}
R_W = \frac{\chi_d/\left(\frac{1}{2}(g\mu_B)^2\rho_0\right)}{\delta\rho/\rho_0}
= 1+\frac{2\rho_0\phi^a}{\alpha} =2,
\end{equation}
we have one parameter
\begin{equation}
\alpha = 2\rho_0 \phi^a.
\end{equation}
Although we assumed that $\alpha$ and $\phi^a$ are small, if we can continue them to
the strong coupling region assuming the Fermi liquid, we can set $\alpha\simeq 1/T_K$ 
which is the only one parameter
of the s-d system.

The Fermi liquid state of the Kondo system can be realized on the basis of the
Anderson model\cite{and61} based on the perturbation theory in the symmetric case with the
condition $E_d=-\frac{1}{2}U$ where $E_d$ is the $d$-electron level and $U$ is the
repulsive interaction between localized $d$ electrons.\cite{yos70,yam75,yos75}
The physical quantities are parametrized by two parameters $\gamma$ and 
$\chi_{\uparrow\downarrow}$.  The Wilson ratio becomes 2 in the limit of large $U$.
These Fermi liquid properties are consistent with the exact solution of the
Anderson model.\cite{kaw81,kaw82}

\section{Kondo effect to Kondo physics}

\subsection{Heavy electrons}

A material in heavy electron systems contains magnetic impurities on each site and
is often realized in rare earth compounds.
The 4$f$ electrons in rare earth atoms play a role of localized electrons to form
the dense Kondo system.
The heavy electron materials exhibit many interesting properties such as heavy effective
electron mass, unconventional superconductivity, unusual magnetic structures and
also topological insulating states.  

Here we discuss briefly the heavy mass realized in a heavy electron material.
The model for heavy electrons is usually the Kondo lattice model or the Anderson
lattice (periodic Anderson) model.
We employ the Hamiltonian given by
\begin{equation}
H= -t\sum_{\langle ij\rangle\sigma}c_{i\sigma}^{\dag}c_{j\sigma}
+V\sum_{i\sigma}(c_{i\sigma}^{\dag}f_{i\sigma}+f_{i\sigma}^{\dag}c_{i\sigma})
+E_f\sum_{i\sigma}f_{i\sigma}^{\dag}f_{i\sigma}
+U\sum_{i}n_{fi\uparrow}n_{fi\downarrow},
\end{equation}
where $c_{i\sigma}$ ($c_{i\sigma}^{\dag}$) and $f_{i\sigma}$ ($f_{i\sigma}^{\dag}$) 
denote annihilation (creation) operators of conduction and localized electrons, 
respectively, and $n_{fi\sigma}=f_{fi\sigma}^{\dag}f_{i\sigma}$.

The strong electron correlation plays an significant role in the realization of heavy electron
states.  We show the result obtained by the optimized variational Monte Carlo method
where the wave function is given in the form
\begin{equation}
\psi = e^{-\lambda K}P_G\psi_0.
\end{equation}
$\psi_0$ is the non-interacting wave function for $U=0$ and $P_G$ is the Gutzwiller operator
for $f$ electrons given as $P_G= \prod (1-(1-g)n_{fi\uparrow}N_{fi\downarrow})$.
$K$ denotes an operator to optimize the wave function.  We take, for example, $K$ to be the 
kinetic energy part of the Hamiltonian.
$g$ and $\lambda$ are variational parameters in the range of $0\le g\le 1$ and
$\lambda >0$.

We show the momentum distribution function as a function of the wave number in Fig. 4.
This shows the heavy electron state with the effective mass 100 times of the band
mass $m_0$: $m_{{\rm eff}}\simeq 100m_0$ for $E_f=-2$.  There is a small jump at the 
Fermi wave number.
The strong electron correlation results in a heavy electron state.
  
\begin{figure}[ht]
\includegraphics[width=8.0cm]{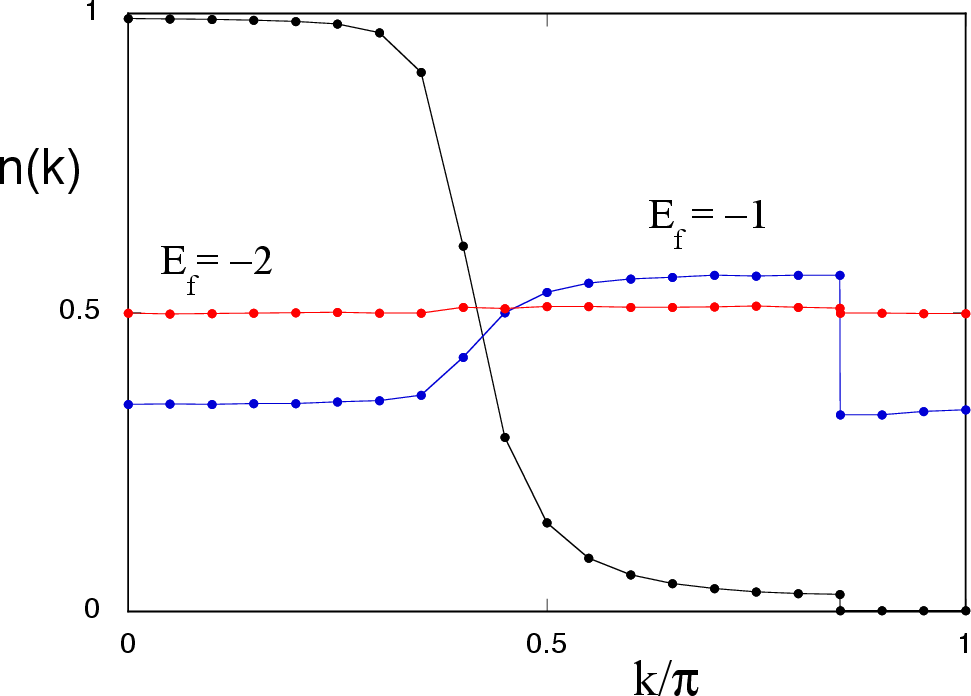} 
\caption{Momentum distribution of the one-dimensional periodic Anderson model.
The calculations are carried out on an $N=40$ lattice with the hybridization $V=0.5t$,
the Coulomb interaction $U=10t$, and the number of electrons is $N_e=70$.
We show the momentum distribution of conduction electrons for $E_f=-1$ (black),
the $f$-electron momentum distribution function for $E_f=-1$ (blue) and that
$E_f=-2$ (red).
We employ the improved wave function in the form $\psi= e^{-\lambda K}\psi_{G}$
where $K$ indicates the kinetic part of the Hamiltonian and $\psi_G$ stands for
the Gutzwiller wave function.\cite{yan98,yan99,yan16b}  The state with $E_f=-2$ represents 
a heavy
mass state with the effective mass $m_{{\rm eff}}/m_0\simeq 100$.  
}
\label{fig4}
\end{figure}

\subsection{Kondo effect and RKKY interaction}

In heavy electron materials the interplay between the Kondo effect and the RKKY interaction
is very important.  This interplay may be crucial in determining the magnetism of heavy
electron states.
As a starting step, the Doniach diagram is often referred to understand the
antiferromagnetism of heavy electrons.\cite{don77}
We discuss the Doniach diagram from the viewpoint of scaling properties of the
s-d exchange coupling and the RKKY interaction.
There may be a transition between an antiferromagnetic state and a non-magnetic Kondo-like 
singlet state as the exchange coupling $J$ is varied.  There may be a critical value $J_c$
at which the transition occurs.  The binding energy of a Kondo singlet $E_K$ and that of
the RKKY antiferromagnetic state $E_{AF}$ are, respectively, given by
\begin{equation}
E_K \simeq De^{-1/\rho J}, ~~~~~ E_{AF}\simeq cD(\rho J)^2,
\end{equation}
where $c$ is a dimensionless constant and $D\sim \rho^{-1}$.  
The Doniach diagram means that by comparing $E_K$ and $E_{AF}$
the qualitative understanding of the magnetic transition may be obtained.

In the language of the renormalization group, the Doniach argument suggests that scaling
equations are given by
\begin{equation}
\mu\frac{\partial J}{\partial\mu}= -\rho J^2,~~~~~\mu\frac{\partial J_{AF}}{\partial\mu}= -J_{AF},
\end{equation}
where $J_{AF}$ stand for the RKKY interaction strength.  From the second equation we have
for $J_{AF}= J_{AF}(\mu)$ as
\begin{equation}
J_{AF}(\mu)\mu = J_{AF}(\mu_0)\mu_0,
\end{equation}
where $\mu_0$ is the initial value of $\mu$ that is set to $D$: $\mu_0= D$.  We set the initial
value of $J_{AF}$ as $J_{AF}^0= J_{AF}(\mu_0)$.  $J_{AF}^0$ is reasonably given by
$J_{AF}^0 = cD(\rho J_0)^2$ (Doniach's $E_{AF}$).
The equation for $J$ results in
\begin{equation}
\rho J(\mu) = \frac{\rho J_0}{1+\rho J_0\ln(\mu/\mu_0)}= \frac{\rho J_0}{1-\rho J_0\ln(J_{AF}(\mu)/J_{AF}^0)}.
\end{equation}
We define the critical value $J_0^c$ of $J_0$ such that when $J_0= J_0^c$, $\rho J$ and $\rho J_{AF}$
increase to be of order one at the same time.  This means
\begin{equation}
1= \frac{\rho J_0^c}{1+\rho J_0^c\ln(\rho J_{AF}^0)}.
\end{equation}
Then we have
\begin{equation}
De^{-1/\rho J_0^c} = e^{-1}D\rho J_{AF}^0 = cD(\rho J_0^c)^2,
\end{equation}
where we used $J_{AF}^0 = cD(\rho J_0)^2$.
This is the relation for the critical value $J_0^c$ that agrees with the Doniach criterion.

Both $J$ and $J_{AF}$ increase as the energy scale $\mu$ decreases from initial values.
Since the scaling equation for $J_{AF}$ is linear in $J_{AF}$, $J_{AF}$ increases faster than $J$
when $J_0$ is small.  Hence the RKKY $J_{AF}$ will be dominant over a Kondo singlet state for small $J_0$.
The Doniach diagram agrees with a scaling picture in the lowest order equation.

\subsection{Fermi surface effect}

Prof. Kondo tried to understand the Kondo effect in a larger framework.\cite{kon87}
He called this framework the Fermi surface effect.
What is the Fermi surface effect?  This question is closely related to the energy
scale of metal electrons.  It is of the order of the Fermi energy $\epsilon_F$ in the
case of static perturbations acting on electrons in metals.  On the other hand,
when the perturbation is dynamical and local, low energy excitation modes come to
play an important role.  These low energy modes give rise to an infrared divergence
that dominates low energy properties.
This effect is called the Fermi surface effect.

The following are examples of the Fermi surface effect:
(1) Kondo effect, (2) X-ray absorption (or emission) of metals,\cite{mah67,mah67b} 
(3) Anderson's orthogonality theorem,\cite{and67,and67b}
(4) Two-level systems in metals.\cite{kon87,kon76,kon84}
(5) Diffusion of heavy particles in metals,\cite{kon87,kon84b,kad86,bre86}

Let us consider a metal with perturbation $V({\bf r})$.  The perturbed wave function
is written as
\begin{equation}
\Psi = N_0\bigg[ \Phi_0+\sum_{|{\bf k}|<k_F, |{\bf k}'|>k_F}
\frac{V_{k-k'}}{\epsilon_{{\bf k}}-\epsilon_{{\bf k}'}}\Phi_0({\bf k}\rightarrow{\bf k}')
+ \cdots\bigg],
\end{equation}
where $\Phi_0$ denotes the Fermi sea and $V_{k-k'}$ is the Fourier transform of
$V({\bf r})$.  $\Phi_0({\bf k}\rightarrow{\bf k}')$ indicates the excited state 
with a hole with momentum ${\bf k}$ in the Fermi sea and an electron with ${\bf k}'$ above 
the Fermi energy.  When $V({\bf r})$ is the local potential, $V_{k-k'}$ is
almost constant.  
From the normalization of the wave function, $N_0$ satisfies
\begin{equation}
1= N_0^2\bigg[ 1+\sum_{|{\bf k}|<k_F, |{\bf k}'|>k_F}
\frac{|V_{k-k'}|^2}{(\epsilon_{{\bf k}}-\epsilon_{{\bf k}'})^2}+ \cdots \bigg].
\end{equation}
When the excitation energy $\epsilon_{{\bf k}}-\epsilon_{{\bf k}'}$ is small, 
$|V_{k-k'}|^2/(\epsilon_{{\bf k}}-\epsilon_{{\bf k}'})^2$ becomes large, then the
summation diverges.  This means $N_0=0$.  This divergence, however, never brings
about a difficulty in the evaluation of expectation values.  The low energy
excitation modes never cause a singularity for the potential $V({\bf r})$.
There is no characteristic energy scale in this case.

On the other hand, in the dynamical problem where 'dynamical' indicates that the
potential changes over time, a divergence could appear in a physical quantity
due to excitation modes of low energy.  This applies to examples of the Fermi
surface effect shown above. 
The importance of the normalization constant $N_0$ was noticed by Anderson.
The inner product of $\Psi$ and $\Phi_0$ gives $\langle\Phi_0|\Psi\rangle=N_0$.
The vanishing of $N_0$ means that the overlap integral also vanishes.
Anderson found that\cite{and67,and67b}
\begin{equation}
\langle\Phi_0|\Psi\rangle = \exp\left( -\frac{1}{2}\left(\frac{\delta}{\pi}\right)^2
\log N_F \right),
\label{overlap}
\end{equation}
where $\delta$ is the $s$-wave phase shift at the Fermi level for the potential, which
is assumed to cause only $s$-wave scattering. $N_F$ is the number of $s$-wave electrons.
The overlap integral in eq.(\ref{overlap}) tends to zero as the system
size increases to infinity.
This is the Anderson orthogonality theorem.  When there are two different local
potentials, the wave functions corresponding to these potentials are orthogonal
each other.
 
The Anderson orthogonality theorem is related to the following overlap integral
\begin{equation}
\langle \Phi_0 e^{iH_0t/\hbar}e^{-i(H_0+V)t/\hbar}\Phi_0\rangle,
\end{equation}
where $H_0$ is the non-interacting Hamiltonian and $V$ is a local potential.
Since $e^{-i(H_0+V)t/\hbar}\Phi_0$ approaches $\Psi$ in the limit $t\rightarrow\infty$,
this overlap integral should vanish in this limit due to the orthogonality theorem.
In the limit $t\gg \hbar/\epsilon_F$, this overlap integral is given 
by\cite{noz69,noz69b,noz69c}
\begin{equation}
\left(\frac{i\epsilon_F t}{\hbar}\right)^{-2(\delta/\pi)^2}.
\end{equation}
Similarly, the integral
\begin{equation}
\langle \Phi_0 e^{iH_0t/\hbar}\sum_kc_k e^{-i(H_0+V)t/\hbar}\sum_kc_k^{\dag}\Phi_0\rangle,
\end{equation}
behaves for large $t$ as,
\begin{equation}
\left(\frac{i\epsilon_F t}{\hbar}\right)^{-1+2\delta/\pi-2(\delta/\pi)^2}.
\end{equation}
This quantity played an important role in deriving the effective Coulomb gas model
from the s-d model.
This behavior appeared to be consistent with the X-ray spectra of metals.\cite{mah81} 

Kondo investigated the muon diffusion in copper.\cite{kon87,kon84,kon84b}  
This issue is reduced to the
evaluation of the overlap integral just like that examined above.
Kondo successfully explained the temperature dependence of the hopping rate of
the positive muon in copper that was reported by muon experiments.
This is also an example of the Fermi surface effect.

\subsection{Quantum dots}

Quantum dots are now where the Kondo effect plays an active role.
A quantum dot is a small puddle of charge containing a well-defined number of
electrons.
In quantum dots a small number of electrons are confined in a finite region of space.
The quantum dot contains a few tens of electrons, in the typical case, and 
is called an artificial atom.
The Kondo effect has been modeled in quantum dots where a quantum dot is connected
by tunnelling junctions to two electron reservoirs through electrode.
These form electron-transport channels known as the Kondo model.
Usually one attaches two leads which are called the source and the drain, and the dot
and leads are connected weakly.
 
We can work out the transmission probability for the transition of an electron
from one lead to the other based on the Anderson model or the Kondo model (s-d model).
We can obtain the s-d interaction in the same way as in the derivation of that
from the Anderson model.  As in the case of magnetic impurities, the transmission
probability is independent of temperature in the lowest order.  As expected,
terms with logarithmic temperature dependence appear at higher orders.
At low temperatures, the transmission probability increases as the temperature
decreases, and approaches unity with temperatures going down to absolute 
zero.\cite{gla88,kaw91,ng88}
This indicates that the conductance tends to $2e^2/h$ at absolute zero.
The physics of quantum dots will continue to make great progress along with
the Kondo effect.

\section{Summary}

We discussed the Kondo physics from several points of view.  The discovery of the Kondo effect by
Prof. Kondo has given a great impact on physics.  He recognized that the resistance minimum
phenomenon is universal for metals with dilute magnetic impurities and he was convinced that 
we could
understand it from a simple and general model.  Thus he investigated the s-d model carefully by
checking various quantum mechanical processes including higher order corrections.
He finally found an unexpected logarithmic term in the resistivity and was convinced that this
term would explain the resistance minimum.  His successful explanation of the resistance minimum
was surprising for physicists in the world.

\begin{acknowledgments}
The author expresses his sincere thanks to the organizing
committee of SCES 2023 for giving me an opportunity to talk about the Kondo physics.
Numerical calculations were carried out on Yukawa-21 at Yukawa Institute of Theoretical
Physics in the Kyoto University and the Supercomputer Center of the Institute for Solid
State Physics, the University of Tokyo. 
\end{acknowledgments}

%\begin{references}

%\end{references}

\end{document}